\documentclass[a4paper,11pt]{article}
\usepackage{pos}

\usepackage{tikz-cd} 
\usetikzlibrary{decorations.markings} 
\usetikzlibrary{shapes} 
\usetikzlibrary{positioning, shapes.geometric} 
\usetikzlibrary{arrows, arrows.meta} 
\usetikzlibrary{decorations.pathreplacing,decorations.markings} 
\tikzset{   ->-/.style args={#1 #2 #3}{     decoration={       markings,         mark=         at position #1         with {\arrow{Stealth[#3,length=#2]}},     },       postaction=decorate, opacity=1.0   },     ->-/.default= 0.5 6pt black } 

\newcommand{\boundellipse}[3]
{(#1) ellipse (#2 and #3)
} 
\definecolor{ao}{rgb}{0.0, 0.5, 0.0} 
\usepackage[framemethod=tikz]{mdframed}

\title{Study of a lattice 2-group gauge model}

\author[a]{Arkadiusz Bochniak}
\author[a]{Leszek Hadasz}
\author*[a]{Piotr Korcyl}
\author*[a]{Błażej Ruba}

\affiliation[a]{Institute of Theoretical Physics, Jagiellonian University\\
prof.  {\L}ojasiewicza 11, 30-348 Krak{\'o}w, Poland}

\emailAdd{arkadiusz.bochniak@doctoral.uj.edu.pl}
\emailAdd{leszek.hadasz@uj.edu.pl}
\emailAdd{piotr.korcyl@uj.edu.pl}
\emailAdd{blazej.ruba@doctoral.uj.edu.pl}

\abstract{Gauge theories admit a generalisation in which the gauge group is replaced by a finer algebraic structure, known as a 2-group. The first model of this type is a Topological Quantum Field Theory introduced by Yetter. We discuss a common generalisation of both the Yetter’s model and Yang-Mills theory and in particular we focus on the lattice formulation of such model for finite 2-groups.
In the second part we present a particular realization based on a 2-group constructed from $\mathbb Z_4$ groups. In the selected model, independent degrees of freedom are associated to both links and faces of a four-dimensional lattice and are subject to a certain constraint. We present the details of this construction, discuss the expected dynamics in different regions of phase space and show numerical results from Monte Carlo simulations corroborating these expectations.}

\FullConference{%
 The 38th International Symposium on Lattice Field Theory, LATTICE2021
  26th-30th July, 2021
  Zoom/Gather@Massachusetts Institute of Technology
}

\begin{document}
\maketitle

\section{Introduction}

Higher gauge theories are physical models whose degrees of freedom associate definite values to $p$-dimensional geometric objects (possibly for several different $p$). In the conventional case of  $p=1$ these are the Wilson lines. A textbook example \cite{Weinberg} is the $p$-form electrodynamics \cite{pform}. Its basic dynamical field $A$ is a $p$-form and hence can be integrated over $p$-dimensional ``surfaces'' $\sigma$. The field $A$ is subject to gauge transformations $A \mapsto A + d \lambda$ with arbitrary $(p-1)$-form $\lambda$. The integral $\int_\sigma A$ is invariant only if $\sigma$ is closed; otherwise it picks up a boundary term $\int_{\partial \sigma} \lambda$. If $\sigma$ itself is the boundary of a $(p+1)$-dimensional $\Sigma$, then one has an integral representation with manifestly gauge-invariant integrand, $\int_\sigma A = \int_{\Sigma} d A$. There exists a lattice formulation of models of this type in which basic degreees of freedom are localized on lattice $p$-cells (e.g. edges for $p=1$, plaquettes for $p=2$ etc.). Of course several fields corresponding to different $p$ may be included in a single model.

The model recalled above generalizes the abelian gauge theory. Given the success of the Yang-Mills theory in particle physics, it is natural to ask whether there exist non-abelian higher gauge theories. This was answered in the negative already in \cite{pform}. Intuitively this may be understood as follows. Consider a path $\gamma$. Given a general $p=1$ non-abelian field $A$ the corresponding observable is the path-ordered exponential $P e^{i \int_\gamma A}$. If $\gamma$ is partitioned into $n$ consecutive pieces $\gamma_1, \dots , \gamma_n$, then this may be rewritten as $P e^{i \int_{\gamma_n} A} \cdots P e^{i \int_{\gamma_1} A}$. This formula reflects the correspondence between concatenation of paths and multiplication in the underlying gauge group. Ordering of this non-commutative multiplication is dictated by the flow of the fictitious time parametrizing $\gamma$. There is no such natural ordering on surfaces or higher dimensional manifolds. 

There exist algebraic objects more intricate than groups whose structure is meant to capture higher dimensional geometry \cite{baez, pfeiffer, baez_huerta}. In the case considered here those are the $2$-groups \cite{baez1} or, equivalently, crossed modules of groups \cite{brown}. Here we prefer the latter description. They can be used as a generalization of the concept of a gauge group. The corresponding gauge theories involve both $p=1$ and $p=2$ degrees of freedom, which are related by a constraint called fake flatness. Despite the presence of the constraint, it is typically not possible to express one degrees of freedom in terms of the other. Basic observables are associated to loops (e.g. boundaries of lattice plaquettes) and closed surfaces (e.g. boundaries of lattice cubes). It turns out that the value of any closed surface observable always belongs to a certain abelian group, but it is constructed using in general non-commutative multiplications involving degrees of freedom of both types.

There are several motivations to study higher gauge theories. They provide interesting examples \cite{kapth, yetter,porter, martins_porter, gpp08, williamson} of Topological Quantum Field Theories (TQFTs) \cite{atiyah, witten}, and hence are expected to describe certain gapped phases of many body quantum systems. In particular, it has been suggested \cite{gukov, KT14} that the deep infrared behaviour of certain conventional gauge theories may be governed by a topological higher gauge theory. Symmetry Protected Topological (SPT) phases protected by higher symmetries were proposed in \cite{kapth}. Higher gauge fields are also invoked in string theory \cite{saemann} and in certain approaches to bosonization \cite{bos1,bos2,bos3}.

This paper is concerned with a class of lattice higher gauge theories based on crossed modules of finite groups. In Section \ref{sec:description} we describe the main observables, symmetries, topological charge sectors and expected phases. In order to probe the corresponding symmetry breaking patterns we invoke nonlocal order parameters: Polyakov loops and ``Polyakov surfaces''. Dynamics is studied using both exact and Monte Carlo methods (for the latter we specialize to a specific crossed module). Results of our simulations are presented in Section \ref{sec:numerics}.

We show that if no explicit interaction terms are introduced, $p=1$ and $p=2$ components of the crossed module-valued gauge fields decouple on the level of correlation functions of local gauge invariant observables. This should be contrasted with the ordinary non-abelian Yang-Mills theory, in which already the gauge-invariant kinetic term makes the theory fully interacting. Nevertheless, some intrinsic interaction does manifest itself in the topological structure and in the behaviour of nonlocal order parameters. We pay close attention to these topics. To give just one example supporting the physical relevance of such subtle phenomena, in the Hamiltonian formulation they correspond to ground state degeneracy and long range entanglement of ground states depending on topological features of the spatial geometry. More detailed treatment has been presented in \cite{jhep_paper,bhr}.

\section{Description of studied models} 
\label{sec:description}

\subsection{Crossed modules}

A crossed module of groups consists of two groups $\mathcal E$, $\Phi$, a homomorphism $\Delta : \Phi \to \mathcal E$ ($\Delta(\varphi_1 \varphi_2)= (\Delta \varphi_1) (\Delta \varphi_2)$) and an action $\rhd$ of $\mathcal E$ on $\Phi$ by automorphisms. That is, $\rhd$ is a binary operation $\mathcal E \times \Phi \to \Phi$ satisfying
\begin{equation}
\epsilon \rhd (\varphi_1 \varphi_2) = (\epsilon \rhd \varphi_1)(\epsilon \rhd \varphi_2), \qquad (\epsilon_1 \epsilon_2) \rhd \varphi = \epsilon_1 \rhd (\epsilon_2 \rhd \varphi), \qquad 1_{\mathcal E} \rhd \varphi = \varphi. 
\end{equation}
Here $1_{\mathcal E} \in \mathcal E$ is the neutral element. In addition, one requires the so-called {\em Peiffer identities} to hold:
\begin{equation}
\Delta (\epsilon \rhd \varphi) = \epsilon (\Delta \varphi) \epsilon^{-1}, \qquad (\Delta \varphi_1) \rhd \varphi_2 = \varphi_1 \varphi_2 \varphi_1^{-1}.
\end{equation}
Here we restrict attention to crossed modules of finite groups. In general neither $\mathcal E$ nor $\Phi$ has to be abelian. However, Peiffer identities imply that elements $\varphi \in \Phi$ satisfying $\Delta \varphi = 1_{\mathcal E}$ commute with the whole $\Phi$. Furthermore, the image $\mathrm{im}(\Delta)$ of $\Delta$ is a normal subgroup of $\mathcal E$ and hence one may form the quotient group $\mathrm{coker}(\Delta)=\mathcal E / \mathrm{im}(\Delta)$.

Later on we will consider the following example: $\mathcal E = \Phi = \mathbb Z_4 = \{ 0 ,1 ,2 ,3 \}$ (addition is the group operation, all arithmetic being performed mod $4$) with
\begin{equation}
\Delta(n) = 2n , \qquad m \rhd n = (-1)^m n.
\label{eq:our_xmod}
\end{equation}
In this crossed module, groups $\mathcal E$ and $\Phi$ are abelian, but they are nontrivially intertwined by operations $\Delta$, $\rhd$. All simulations are performed for a gauge theory based on this crossed module. On the other hand, some of our exact results do not depend on this choice.

\subsection{Degrees of freedom and the action}

Before we explain how to put degrees of freedom valued in a crossed module on a lattice, we briefly discuss the geometric setup. Lattice sites are called vertices, with typical symbol $v$. For every edge (link) $e$ we choose a direction (from the source vertex $s(e)$ to the target vertex $t(e)$). For every face (plaquette) $f$ we choose a corner $b(f)$ and an orientation. Then the boundary $\partial f$ of $f$ is written as the ``composition'' of some number of edges, starting and ending at $b(f)$. If traversing $\partial f$ one encounters an edge $e$ whose given orientation does not agree with that of $\partial f$, one uses $e^{-1}$ instead.

By a field configuration we shall mean an assignment of $\epsilon_e \in \mathcal E$ to every edge $e$ and $\varphi_f \in \Phi$ to every face $f$ such that the {\em fake flatness} constraint is satisfied:
\begin{equation}
\Delta \varphi_f = \epsilon_{\partial f}.
\end{equation}
Here $\epsilon_{\partial f}$ is the Wilson loop around $\partial f$ built from $\epsilon$ degrees of freedom. For example if $\partial f = e_3 e_2 e_1$ (with $s(e_1) = t(e_3) = b(f)$, $t(e_1)=s(e_2)$ and $t(e_2) = s(e_3)$), then $\epsilon_{\partial f} = \epsilon_{e_3} \epsilon_{e_2} \epsilon_{e_1}$.

Gauge field $\epsilon$ is subject to standard gauge transformations: 
\begin{equation}
\epsilon_e \mapsto \xi_{t(e)} \epsilon_e \xi_{s(e)}^{-1}, \qquad \varphi_f \mapsto \xi_{b(f)} \rhd \varphi_f.
\label{eq:v_transf}
\end{equation}
Note that $\varphi_f$ behaves as a matter field placed at $b(f)$. In addition, one considers also higher transformations, parametrized by group elements $\psi_e \in \Phi$ on edges:
\begin{equation}
\epsilon_e \mapsto (\Delta \psi_e) \epsilon_e, \qquad \varphi_f \mapsto \psi_{e_n} (\epsilon_{e_n} \rhd \psi_{e_{n-1}}) \cdots (\epsilon_{e_n} \cdots \epsilon_{e_2} \rhd \psi_{e_1}) \varphi_f,
\label{eq:edge_transf}
\end{equation}
where we wrote $\partial f = e_n \cdots e_1$.

General transformations \eqref{eq:edge_transf} have the property that Wilson loops are not invariant, which eliminates all local gauge invariant observables built of the $\epsilon$ field. Our objective being to generalize Yang-Mills theory, we regard this as undesirable. There exists a way out by declaring that only transformations \eqref{eq:edge_transf} with $\Delta \psi_e = 1_{\Phi}$ are admitted as gauge transformations. General transformations \eqref{eq:edge_transf} are then used in the analytic study of the system and as local constraint-preserving moves in Monte Carlo simulations. For those purposes we introduce also another class of transformations, parametrized by $\chi_f \in \ker(\Delta)$:
\begin{equation}
    \epsilon_e \mapsto \epsilon_e, \qquad \varphi_f \mapsto \chi_f \varphi_f.
    \label{eq:f_transf}
\end{equation}

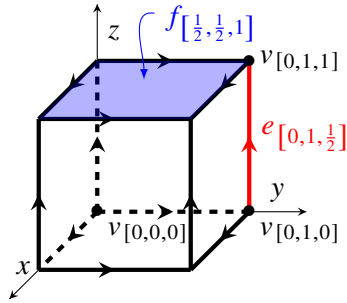
\begin{figure}[ht]
\centering
\begin{tikzpicture}[scale=0.5]
  \draw[ultra thick, dashed, decoration={markings, mark=at position 0.5 with {\arrow{stealth}}},postaction={decorate}]
  (0,0,0) coordinate (O) -- (4,0,0) coordinate (A1);
  \draw[ultra thick, dashed, decoration={markings, mark=at position 0.5 with {\arrow{stealth}}},postaction={decorate}]
  (O) -- (0,4,0) coordinate (A2);
  \draw[ultra thick, dashed, decoration={markings, mark=at position 0.5 with {\arrow{stealth}}},postaction={decorate}]
  (O) -- (0,0,4) coordinate (A3);
\draw[ultra thick, decoration={markings, mark=at position 0.5 with {\arrow{stealth}}},postaction={decorate},color=red]
  (A1) -- node[right] {{\color{red}$e_{\left[0,1,\frac{1}{2}\right]}$}} (4,4,0) coordinate (A4);
\draw[ultra thick, decoration={markings, mark=at position 0.5 with {\arrow{stealth}}},postaction={decorate}]
  (A1) -- (4,0,4) coordinate (A5);
  \draw[ultra thick, decoration={markings, mark=at position 0.5 with {\arrow{stealth}}},postaction={decorate}]
  (A2) -- (0,4,4) coordinate (A6);
\draw[ultra thick, decoration={markings, mark=at position 0.5 with {\arrow{stealth}}},postaction={decorate}]
  (A3) -- (A6);
\draw[ultra thick, decoration={markings, mark=at position 0.5 with {\arrow{stealth}}},postaction={decorate}]
  (A2) -- (A4);
  \draw[ultra thick, decoration={markings, mark=at position 0.5 with {\arrow{stealth}}},postaction={decorate}]
  (A3) -- (A5);
  \draw[ultra thick, decoration={markings, mark=at position 0.5 with {\arrow{stealth}}},postaction={decorate}]
  (A4) -- (4,4,4) coordinate (A7);
  \draw[ultra thick, decoration={markings, mark=at position 0.5 with {\arrow{stealth}}},postaction={decorate}]
  (A5) -- (A7);
  \draw[ultra thick, decoration={markings, mark=at position 0.5 with {\arrow{stealth}}},postaction={decorate}]
  (A6) -- (A7);
  \draw [decoration={markings, mark=at position 1.0 with {\arrow{stealth}}},postaction={decorate}] (A1) -- +(1.5pt,0,0) node [midway,above] {$y$};
  \draw [decoration={markings, mark=at position 1.0 with {\arrow{stealth}}},postaction={decorate}] (A2) -- +(0,1.5pt,0) node [midway,right] {$z$};
  \draw [decoration={markings, mark=at position 1.0 with {\arrow{stealth}}},postaction={decorate}] (A3) -- +(0,0,2pt) node [midway,above] {$x$};
  \node at (O) {$\bullet$};
  \node[below right] at (O) {$v_{\left[0,0,0\right]}$};
  \node at (A1) {$\bullet$};
  \node[below right] at (A1) {$v_{\left[0,1,0\right]}$};
  \node at (A4) {$\bullet$};
  \node[right] at (A4) {$v_{\left[0,1,1\right]}$};

\node (f1) at (3,5,0) {{\color{blue}$f_{\left[\frac{1}{2},\frac{1}{2},1\right]}$}};
\node (f1a) at (2,4,2) {};
\draw[-latex,blue] (f1) to[out=170,in=90] node[midway,font=\scriptsize,above] {} (f1a);
\fill[blue,opacity=0.3] (A2) -- (A6) -- (A7) -- (A4) -- cycle;

\end{tikzpicture}
\caption{The unit cell of the cubic lattice.}
\label{fig:cube}
\end{figure}

Besides Wilson loops, there are also surface observables. For the general formalism needed to write down the corresponding formulas, see \cite{bullivant20,bhr}. Here we confine ourselves to the case of the boundary of an elementary cube in a cubic lattice. Labeling vertices, edges and faces by their midpoint (see Fig. \ref{fig:cube}) and choosing orientations determined by the lexicographic ordering, we have
\begin{align}
\varphi_{\mathrm{cube}} = & \varphi_{[\frac12,\frac12,0]} \varphi_{[\frac12,0,\frac12]} (\epsilon_{[0,0,\frac12]}^{-1} \rhd \varphi_{[\frac12,\frac12,0]}) \varphi_{[\frac12,0,\frac12]}^{-1} (\epsilon_{[\frac12,0,0]}^{-1} \rhd \varphi_{[1,\frac12,\frac12]}) \\
& \varphi_{[\frac12,\frac12,0]}^{-1} (\epsilon_{[0,\frac12,0]}^{-1} \rhd \varphi_{[\frac12,1,\frac12]}) \varphi_{[0, \frac12, \frac12]}^{-1} \varphi_{[\frac12, 0, \frac12]}^{-1} \varphi_{[\frac12, \frac12,0]}^{-1}. \notag
\end{align}
It may come as a surprise that there are $10$ elements $\varphi$ in this formula, but note that the last two factors would cancel with the first two if $\Phi$ was assumed to be abelian. More generally, they could be traded for additional $\epsilon$ factors using Peiffer identities and fake flatness. Then the remaining $\varphi$ factors are in 1-1 correspondence with the six faces of the cube. We note that $\varphi_{\mathrm{cube}}$ transforms as
\begin{equation}
\varphi_{\mathrm{cube}} \mapsto \xi_{[0,0,0]} \rhd \varphi_{\mathrm{cube}}
\end{equation}
under standard gauge transformations. It is invariant under all transformations \eqref{eq:edge_transf}. 

We consider action functionals generalizing the Wilson action:
\begin{equation}
S(\epsilon, \varphi) = J_1 \sum_{\mathrm{plaquettes}} f_1 (\epsilon_{\mathrm{plaquette}}) + J_2 \sum_{\mathrm{cubes}} f_2 (\varphi_{\mathrm{cube}}),
\end{equation}
where $J_1, J_2 \geq 0$ are coupling constants and $f_1 : \mathcal E \to \mathbb R$, $f_2 : \Phi \to \mathbb R$ are fixed functions which have unique minima at the corresponding neutral elements. Then $S(\epsilon, \varphi)$ penalizes excitations of any plaquettes or cubes. Precise form of $f_1, f_2$ functions has to be made separately for every crossed module (in general there could be multiple coupling constants hidden within, but our model has only $J_1, J_2$). We do need $f_1(\xi \epsilon \xi^{-1}) = f_1(\epsilon)$ and $f_2(\xi \rhd \varphi) = f_2 (\varphi)$ for gauge invariance. Precise form of the action used in our simulations will be given later in \eqref{eq:our_action}.

\subsection{Topological charges}

Let $\gamma$ be a non-contractible loop (e.g. winding around a single direction in toric geometry). Then the Wilson line $\epsilon_\gamma$ along $\gamma$ is called a Polyakov loop. For general field configurations its value depends on the shape of $\gamma$. Now let $\overline{\epsilon_\gamma} \in \mathrm{coker}(\Delta)$ be the reduction of $\epsilon_\gamma$ modulo the image of~$\Delta$. Using fake flatness one shows that $\overline{\epsilon_\gamma}$ is invariant to deformations of $\gamma$ -- it depends only on its homotopy class. We will call it (or rather its conjugacy class, for gauge invariance) a topological charge. Topological charges are invariant under all local constraint-preserving moves, obstructing ergodicity in Monte Carlo simulations. In a fixed topological charge sector, any two configurations may be connected by a sequence of local moves of three types (\ref{eq:v_transf}, \ref{eq:edge_transf}, \ref{eq:f_transf}) \cite{bhr}.

\subsection{Factorization theorem}

Let $O_1$ be an observable depending only on plaquette observables and $O_2$ an observable depending only on cube observables. Then restricting the partition sum to a fixed topological charge sector, one has factorization $\langle O_1 O_2 \rangle = \langle O_1 \rangle \langle O_2 \rangle$. Furthermore, $\langle O_1 \rangle$ depends only on $J_1$ and $\langle O_2 \rangle$ depends only on $J_2$. This is proven using the fact that $O_1$ is invariant to \eqref{eq:f_transf} and $O_2$ is invariant to \eqref{eq:edge_transf}, see \cite{jhep_paper} for details.

Depending on the choice of the crossed module, there may or may not exist exact symmetries relating different topological charge sectors, implying that the factorization holds also after summing over sectors. In any case, volume scaling analysis shows that averages of local observables become independent on the sector in the thermodynamic limit. Situation is less trivial for nonlocal order parameters. In the case of \eqref{eq:our_xmod}, we have found an observable whose average depends on both $J_1, J_2$ and the topological charge, see \ref{sec:order_parameters}.

\subsection{Symmetries and order parameters}
\label{sec:order_parameters}

Here we focus on the crossed module \eqref{eq:our_xmod} and toric geometry, see \cite{bhr, jhep_paper} for discussions in more general setting. Invoking the terminology of \cite{global}, this model enjoys two higher-form symmetries.

First, there is a $1$-form $\mathbb Z_2$ symmetry resembling the center symmetry from Yang-Mills theory. It may be implemented as follows: choose a direction $\mu$ and a hyperplane perpendicular to $\mu$. Then shift by $2$ the value of every link variable in the direction $\mu$ and contained in the chosen hyperplane. This preserves all local gauge invariant observables (and hence the action), but it changes the value of Polyakov loops in direction $\mu$. Note that this construction yields as many $\mathbb Z_2$ symmetries as directions, and in general the number of symmetries would depend on the topology. Superficially these $\mathbb Z_2$ transformations depend also on the choice of a hyperplane, but this is a gauge artifact. 

The $2$-form symmetry is a straightforward generalization: one chooses two directions $\mu, \nu$ and shifts by $2$ the value of all plaquette variables in the $\mu \nu$ directions contained in a chosen plane orthogonal to directions $\mu , \nu$.

The Polyakov loop is an order parameter for the $1$-form symmetry. However, in a finite volume system the average of a single Polyakov loop is always trivial, as dictated by the symmetry. Therefore, instead of a single Polyakov loop in the direction $\mu$ we consider the absolute value of the average over the volume perpendicular to $\mu$.

To obtain an order parameter for the $2$-form symmetry (up to suitable averaging), morally speaking one has to add all plaquette variables ($\varphi$, not the plaquettes built of $\epsilon$!) in a $2$-plane parallel to directions $\mu, \nu$. This naive prescription does not yield a gauge-invariant quantity, though. To fix this issue it is necessary to use link variables to parallel transport all plaquettes to a single point. This observable will be called a~Polyakov surface. Similarly as for the Polyakov loop, in a finite volume numerical simulation we will consider averages of absolute value of the Polyakov surface.

\subsection{Phase diagram and relation to TQFTs}

Due to the factorization phenomenon, the phase diagram may be discussed separately as a~function of $J_1$ and $J_2$, as long as we restrict our attention to local observables. 

The fake flatness constraint enforces that all $\epsilon_{\partial f}$ plaquettes lie in $\mathrm{im}(\Delta)$. Based on this observation one may show \cite{jhep_paper} that averages of functions of $\{ \epsilon_{\partial f} \}$ reduce to averages in standard lattice gauge theory with gauge group $\mathrm{im}(\Delta)$. In the case of the crossed module \eqref{eq:our_xmod} one has $\mathrm{im}(\Delta)=\{ 0 , 2 \} \cong \mathbb Z_2$. We are most interested in the dimension $4$, in which the $\mathbb Z_2$ gauge theory enjoys Krammers-Wannier self-duality \cite{wegner}. Due to the duality, location of the (first order) phase transition \cite{creutz_Z2} is known exactly to be $J_1^{\mathrm{crit}} = \frac12 \mathrm{arsinh}(1) \approx 0.441$. In the regime $J_1 < J_1^{\mathrm{crit}}$ averages of plaquettes are small and the $1$-form symmetry is unbroken. For $J_1 > J_1^{\mathrm{crit}}$ plaquette averages are larger and the symmetry becomes broken. We remark that the factorization theorem does apply here because the ratio of two Polyakov loops, in the same direction but displaced, is expressible in terms of plaquettes within the strip enclosed by the two lines.

In the analysis of cube observables, we may use factorization and put $J_1 = \infty$. Then, up to effects vanishing in the thermodynamic limit, $\epsilon$ fields are turned off. The fake flatness constraint enforces all $\varphi_f$ to be valued in the group $\ker(\Delta)$. Hence we recover a $2$-form gauge theory for the abelian group $\ker( \Delta)$, equal to $\mathbb Z_2$ in the case \eqref{eq:our_xmod}. In the dimension $4$, such gauge theory is Krammer-Wannier dual to the Ising model. From the corresponding literature \cite{lundow} we obtain a~prediction for the location of (a second order) phase transition: $J_2^{\mathrm{crit}} \approx 0.953294(1)$. Averages of cubes are small for small $J_2$ and large for large $J_2$. Status of the symmetry is less transparent, though, since the corresponding order parameter, the Polyakov surface defined in Section \ref{sec:order_parameters}, does not obey the factorization theorem. This question is largely addressed by our simulations, see Section \ref{sec:numerics}.

As outlined above, at least for the crossed module \eqref{eq:our_xmod} and dimension $4$ there are $4$ phases determined by two binary choices: $J_1 \lessgtr J_1^{\mathrm{crit}}$ and $J_2 \lessgtr J_2^{\mathrm{crit}}$. They may be thought of as basins of attraction of four renormalization group fixed points with $J_1 \in \{ 0 , \infty \}$, $J_2 \in \{ 0 , \infty \}$. Each of these points may be interpreted as a particular TQFT:
\begin{itemize}
    \item $(J_1, J_2)=(0,0)$: topological $\mathbb Z_2$ gauge theory \cite{Dijkgraaf} of topological charges.
    \item $(J_1 , J_2) = (\infty,0)$: topological $\mathbb Z_4$ gauge theory of Polyakov loops.
    \item $(J_1, J_2) = (0, \infty)$: Yetter's TQFT. One has invariance to all transformations \eqref{eq:edge_transf} and all cube observables are frozen. Neverthless, there remains a rich topological content. In contrast to standard topological gauge theories, it is sensitive to topological invariants of the underlying geometry other than the fundamental group. There is an interplay between line operators (such as topological charges) and surface operators (such as Polyakov surfaces), see Fig. \ref{fig:sketch}.
    \item $(J_1, J_2) = (\infty, \infty)$: topological $\mathbb Z_4$ gauge theory of Polyakov loops and $2$-form $\mathbb Z_2$ gauge theory of Polyakov surfaces, completely independent of each other. We remark that in the analogous limit for general crossed modules, one does not necessarily have factorization on the level of the topological structure (compare with \cite[Sec. 3.4]{bhr}).
\end{itemize}

In \cite{bhr} a Hamiltonian version of our model has been considered and four integrable Hamiltonians corresponding to the limits above have been constructed.

\section{Numerical simulations}
\label{sec:numerics}
\subsection{Notation}

In this section we present results obtained through numerical Monte Carlo simulations of the gauge theory based on the crossed module \eqref{eq:our_xmod}, defined on a four-dimensional lattice with periodic boundary conditions. For further convenience we denote the lattice extent in direction $\mu$ by $L_\mu$; we performed our simulations with $L_0=L_1 = L_2 \equiv L$, often with different $L_3$.


For a large part of the preceding discussions, fairly general geometries are admissible. Restriction to cubic lattices allows to introduce simpler notations. Lattice sites are labeled by $\mathbf x = (x,y,z,t)$. We denote by $\widehat \mu$ the unit vector in $\mu$-th direction. Let $m_{\mu}(\mathbf x)$ be the variable associated to a link in direction $\mu$, starting at lattice site $\mathbf x$. Similarly, for plaquette variables in $\mu \nu$ plane we write $n_{\mu \nu}(\mathbf x)$. We also have plaquettes $f_{\mu \nu}$ built of links in standard way; they appear in the fake flatness constaint
\begin{equation}
    2 n_{\mu \nu}(\mathbf x)=f_{\mu \nu}(\mathbf x),
\end{equation}
which only partially determines $n_{\mu \nu}$ in terms of links. Observables corresponding to cubes with edges in directions $\mu, \nu, \rho$ will be denoted by $g_{\mu \nu \rho}(x,y,z,t)$. We chose the following action:
\begin{equation}
    S = J_1 \sum_{\mathbf x} \sum_{\mu < \nu} (-1)^{\frac{f_{\mu \nu}(\mathbf x)}{2}} + J_2 \sum_{\mathbf x} \sum_{\mu < \nu < \rho} (-1)^{\frac{g_{\mu \nu \rho}(\mathbf x)}{2}}.
    \label{eq:our_action}
\end{equation}


\subsection{Algorithm}

The numerical simulation is based on the Metropolis algorithm \cite{metropolis1,metropolis2} where local update movements, separate for the link and faces variables, have been modified in order to accommodate the fake-flatness constraint. Hence, the constraint is preserved by the updates, i.e. if it was satisfied by the local degrees of freedom in the initial configuration, it remains fulfilled during the entire simulation. Additional updates preserving the system energy were sparsely incorporated in between local update movements in order to decrease autocorrelation times. Their role is to change the sign of all Polyakov lines and all Polyakov planes and play the role of over-relaxation steps as in Refs \cite{overrelaxation1,overrelaxation2,overrelaxation3,overrelaxation4}. In each point of the parameter space we generate Markov chains of length of the order of $10^5$ configurations spaced by at least $ \sim V$ local updates. In order to ensure ergodicity we repeat the simulations with different pseudo-random number generator seed and always compare simulations started with a cold and hot initial configurations. Numerical results presented below come from simulations where the deviations between such additional simulations were smaller than statistical uncertainties. In order to estimate the latter we use the explicitly calculated autocorrelation function which we integrate up to the first non-positive element to conclude the associated autocorrelation time. All data points shown are presented together with their statistical uncertainties, however in many cases the error bars are smaller than the symbol size, hence may not be fully visible. The algorithm as described above does not allow to jump between configurations with different values of the topological charge. The update which would change the sign of the latter is non-local and leads to a very poor acceptance rate. We therefore consider separate simulations at fixed topological charge.

In order to study the phase space of our system we monitor a number of observables. They can be classified into two classes: local and non-local observables. We define them separately in the following two subsections and discuss the dynamics of the system unveiled by their numerical estimates.

\subsection{Local observables and phase transitions}

\begin{figure}
    \centering
    \includegraphics[width=0.32\textwidth, angle=270]{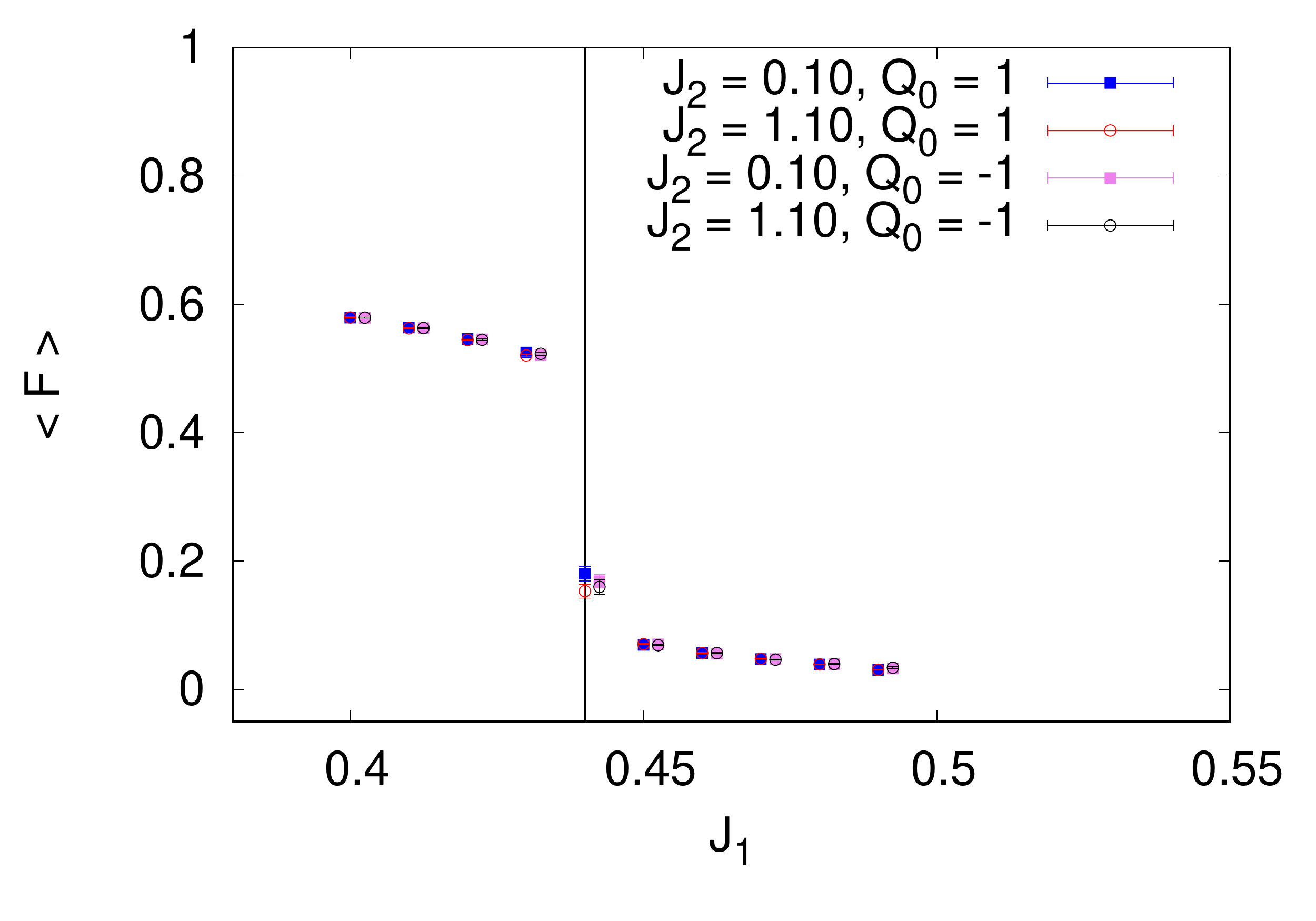}
    \includegraphics[width=0.32\textwidth, angle=270]{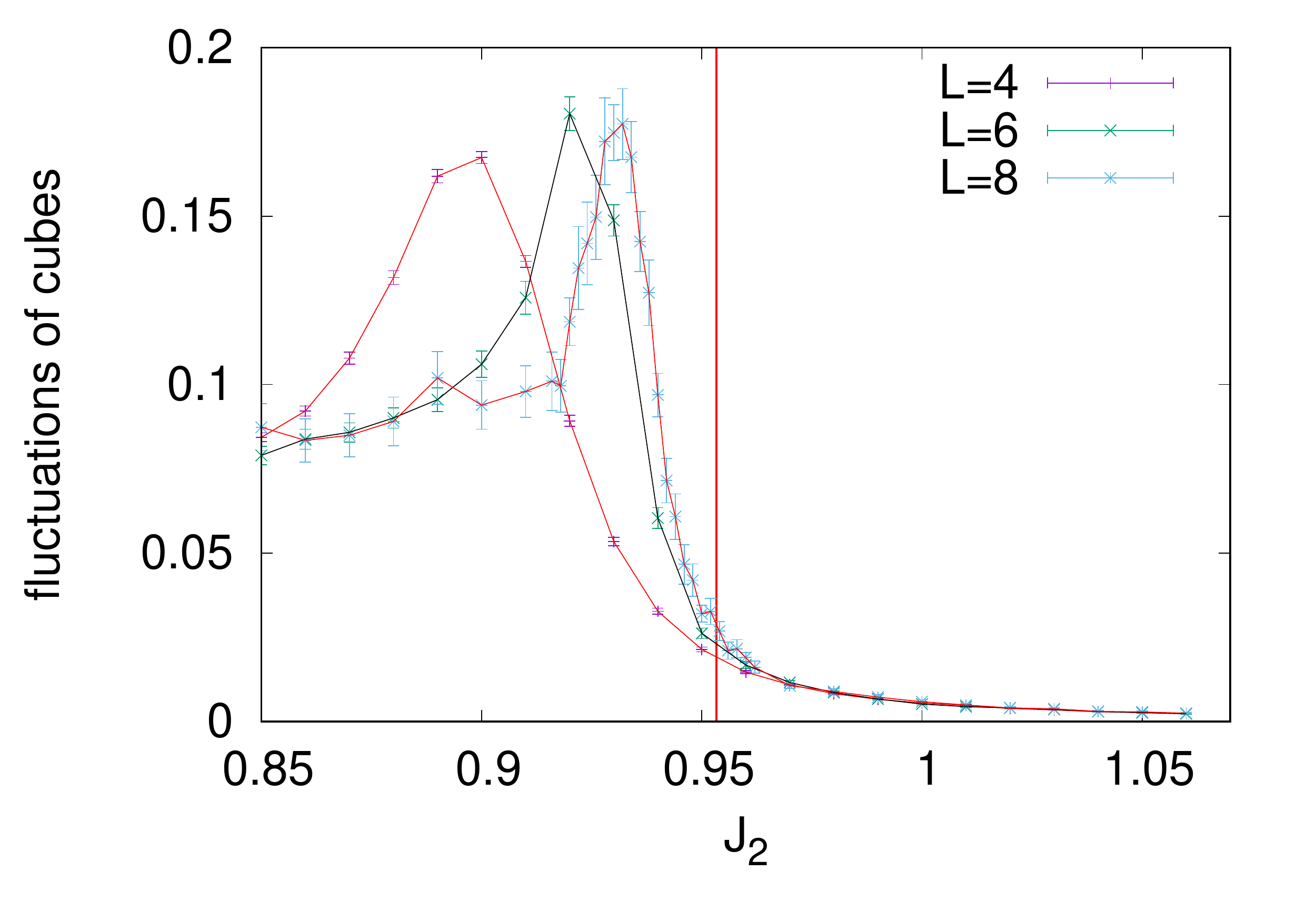}
    \caption{Estimates of the average plaquette Eq.\ref{eq:average plaquette} and average cube Eq.\ref{eq:average cubes} as a function of appropriate coupling constants $J_1$ and $J_2$.} 
    \label{fig:local}
\end{figure}

The simplest observables which we monitor as we sweep over the parameter space are the quantities building the action of the system \eqref{eq:our_action}. Hence, let us define the average value of the plaquettes $F$ and cubes $G$ as
\begin{align}
    F &=  \Big| \frac{1}{6 V} \sum_{x,y,z,t} \sum_{\mu < \nu} f_{\mu \nu}(x,y,z,t) \Big|, \label{eq:average plaquette}\\
    G &=  \Big| \frac{1}{4 V} \sum_{x,y,z,t} \sum_{\mu < \nu < \rho} g_{\mu \nu \rho}(x,y,z,t) \Big|. \label{eq:average cubes}
\end{align}
Here $V$ is the total number of lattice sites.

In Figure \ref{fig:local} we present their dependence on the appropriate coupling constant. Note that due to the factorization theorem \cite{jhep_paper} the expectation values of such local operators can depend only on one of the coupling constant. The left panel of Figure \ref{fig:local} indeed provides confirmation of such behavior. Namely, the expectation values of the plaquette estimated at two very different values of the $J_2$ coupling constant are always compatible within their statistical uncertainties and exhibit a sharp drop in value around the expected first order phase transition at $J_1 \approx 0.44$. This is also true irrespective of the topological sector as we are simultaneously showing data coming from two sectors with different topological charge. The right panel of Figure \ref{fig:local} shows the fluctuations of the $\langle G \rangle$ observable as a function of the $J_2$ coupling constant for estimated using three lattice volumes. We notice a sharp maximum in that observable which approaches with increasing volume the position of the expected second order phase transition in the $J_2$ coupling constant. These two illustrative examples provide numerical evidence of two main facts. First, that the system has four distinctive regions on phase space separated by two critical values of the couplings $J_1$ and $J_2$. Second, that the factorisation theorem holds, and therefore also the dynamics as seen through the local observables such as $\langle F \rangle$ and $\langle G \rangle$ factorizes and we do not see any interplay between the dynamics of the links and faces. 

\subsection{Non-local observables}

In this subsection we turn our attention to non-local observables. We investigate two such observables: $P_{\mu}$ and $P_{\mu \nu}$, which we call Polyakov line and plane respectively. Here we just state the final formulae and refer the Reader to \cite{jhep_paper} for more details:
\begin{equation}
 P_{\mu} = \left| V_{\perp}^{-1} \sum_{\mathbf x} \exp \left(\frac{i \pi}{2}{\sum \limits_{j=0}^{L_{\mu}-1} m_{\mu}(\mathbf x + j \widehat \mu)}\right) \right|,
 \label{eq:pmu_avg}
\end{equation}
with the sum taken over $x$ in a plane transverse to the $\mu$-th direction and $V_{\perp} = \prod \limits_{\nu \neq \mu} L_{\nu}$ is the transverse volume. Topological charges take the form
\begin{equation}
Q_{\mu} = \exp \left( i \pi \sum \limits_{j=0}^{L_{\mu}-1} m_{\mu}(\mathbf x + j \widehat \mu) \right),
\end{equation}
which does not depend on $\mathbf x$ by fake flatness. Thirdly, we need the Polyakov planes:
\begin{equation}
P_{\mu \nu} = \left| V_{\perp}^{-1} \sum_x \exp \left( \frac{i \pi}{2} \Sigma_{\mu \nu}(x) \right) \right|,
\label{eq:pmunu_avg}
\end{equation}
where $V_{\perp} = \prod \limits_{\rho \neq \mu, \nu} L_{\rho}$ and the sum is taken over a plane transverse to $ \mu$ and $ \nu$. $\Sigma_{\mu \nu}(x)$ is the sum of all $n_{\mu \nu}(x)$ in a plane winding around two lattice directions $\hat{\mu}$ and $\hat{\nu}$, with appropriate link factors included to ensure gauge invariance.

%

\begin{figure}
    \centering
    \includegraphics[width=0.32\textwidth]{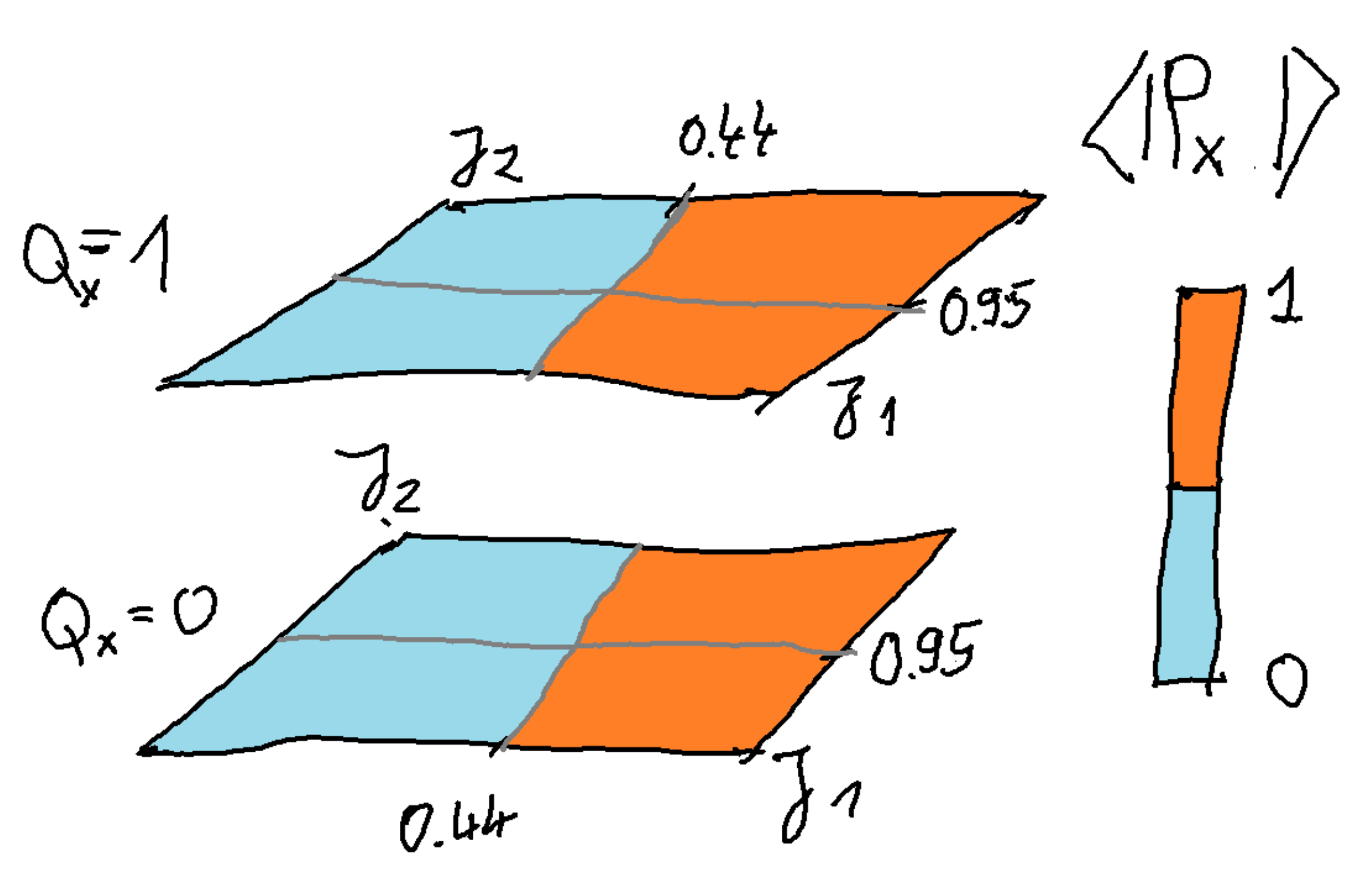}
    \hspace{2cm}
    \includegraphics[width=0.32\textwidth]{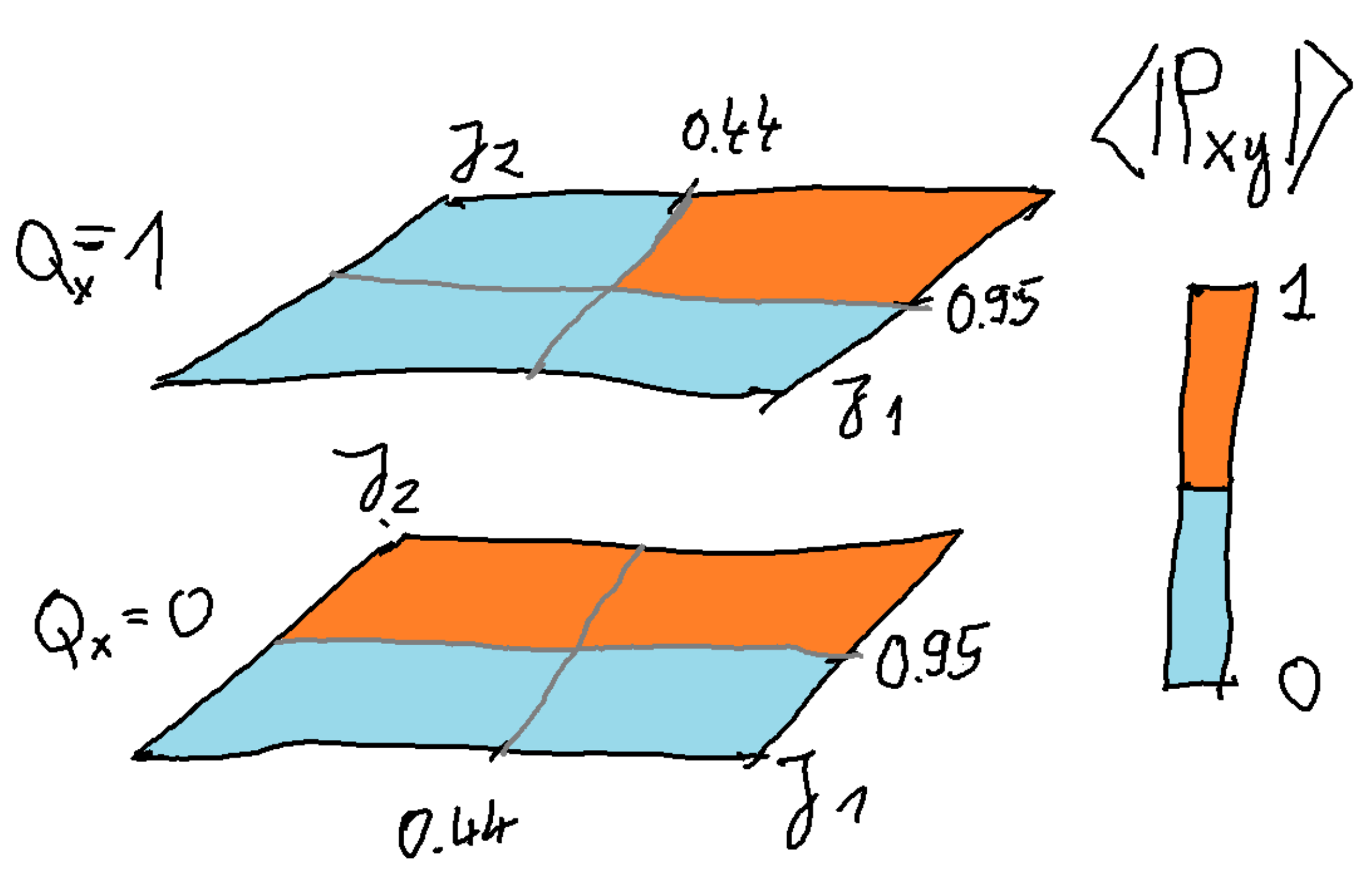}
    \caption{Sketch showing the magnitude of the Polyakov line $\langle P_0 \rangle$ and Polyakov surface $\langle P_{01} \rangle$ as a function of $J_1$, $J_2$ and topological charge. Precise values of these observables at the four points in phase space are contained in Table \ref{tab:polyakov plane values}.  Approximately values of the critical couplings are shown.}
    \label{fig:sketch}
\end{figure}

We start the discussion of numerical results for $P_{\mu}$ and $P_{\mu \nu}$ with the presentation of the expectation values of $P_{01}$ in the four regions of phase space collected in Table \ref{tab:polyakov plane values}. A schematic view of the overall situation is shown in Figure \ref{fig:sketch}. The particular values of $J_1$ and $J_2$ have been selected based on the observations of phase transitions exhibited by local observables $F$ and $G$ and are located such that there is one pair of coupling constants per phase. Moreover, we include results for two topological sectors, with charge $Q_{\mu}=1$ and $Q_{\mu}=-1$ in the $\mu=0$ direction. The first thing to notice is that the factorization theorem does not hold anymore and $\langle P_{01} \rangle$ has a nontrivial dependence on both $J_1$ and $J_2$ coupling constants and on the topological charge. This demonstrates that the fake-flatness constraint on one hand indeed modifies significantly the dynamics, and on the second hand still allows for independent dynamics in the two kind of dynamical degrees of freedom, namely links and faces. Second thing to notice is that the topological charge plays a crucial role. In the sector with $Q_0=1$, the dynamics of $P_{01}$ depends only on the $J_2$ coupling constant, which can be seen by comparing the first two rows and the last two rows in the left part of Table \ref{tab:polyakov plane values}. The most interesting piece of information contained in these tables is located in the last two rows of the table on the right. In the sector with $Q_0=-1$ the Polyakov plane is also sensitive to the phase transition in the $J_1$ coupling constant. Hence, it can be used to monitor the dynamics of the system on the entire plane of parameters.

\begin{table}[]
    \centering
    \begin{tabular}{|c|c|c||c|}
         \hline
         $J_1$ & $J_2$ & $Q_0$ & $\langle P_{01} \rangle$ \\
         \hline
         0.43  & 0.1 & 1  & 0.0631(2)\\
         \hline
         0.46  & 0.1 & 1  & 0.0630(1)\\
         \hline
         0.43  & 1.1 & 1  & 0.9815(3)\\
         \hline
         0.46  & 1.1 & 1  & 0.9838(2)\\
         \hline
    \end{tabular}
    \hspace{1cm}
    \begin{tabular}{|c|c|c||c|}
         \hline
         $J_1$ & $J_2$ & $Q_0$ & $\langle P_{01} \rangle$ \\
         \hline
         0.43  & 0.1 & -1 & 0.0631(2)\\
         \hline
         0.46  & 0.1 & -1 & 0.0630(1)\\
         \hline
         0.43  & 1.1 & -1 & 0.0720(2)\\
         \hline
         0.46  & 1.1 & -1 & 0.9238(1)\\
         \hline
    \end{tabular}
    \caption{Average values of $\langle P_{01}\rangle$ in the four regions of phase diagram estimated on a lattice with $L=4$ and $L_3 = 40$. Data on the left correspond to the $Q_0=1$ topological sector whereas the data on the right to the $Q_0=-1$ sector.}
    \label{tab:polyakov plane values}
\end{table}

\begin{figure}
    \centering
    \includegraphics[width=0.32\textwidth, angle=270]{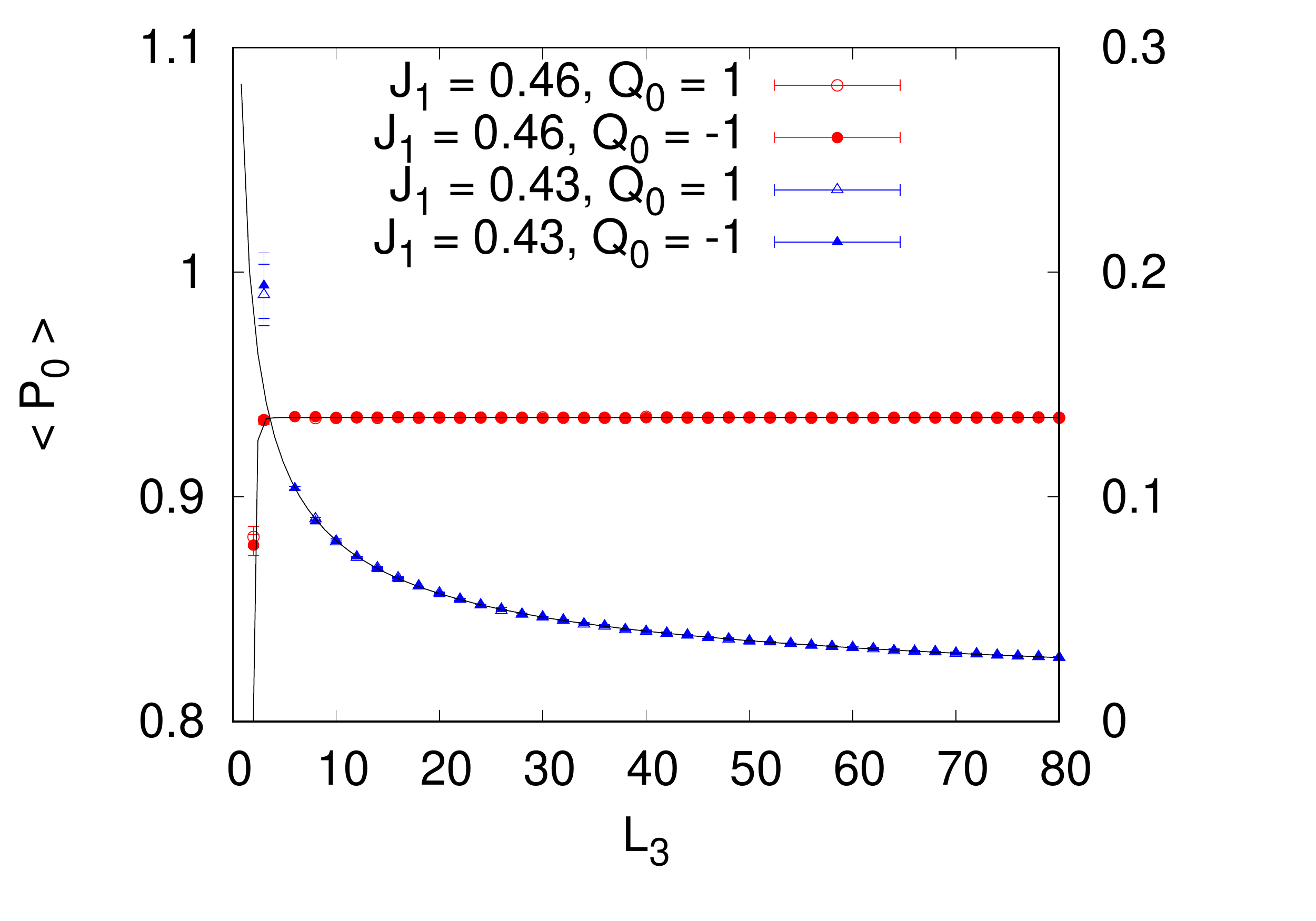}
    \includegraphics[width=0.32\textwidth, angle=270]{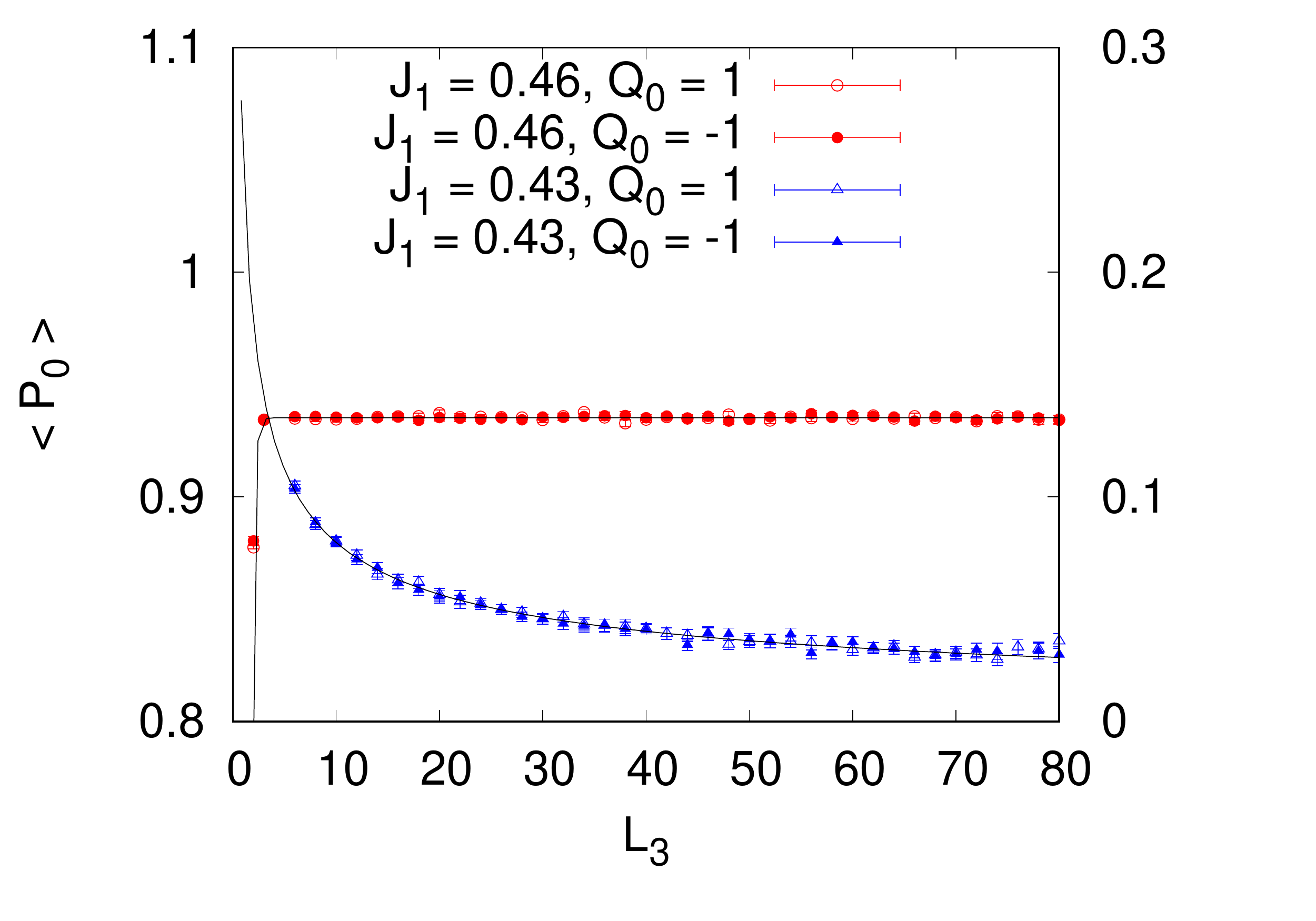}
    \caption{Results for the expectation value of the Polyakov line}
    \label{fig:line}
\end{figure}

\begin{figure}
    \centering
    \includegraphics[width=0.32\textwidth, angle=270]{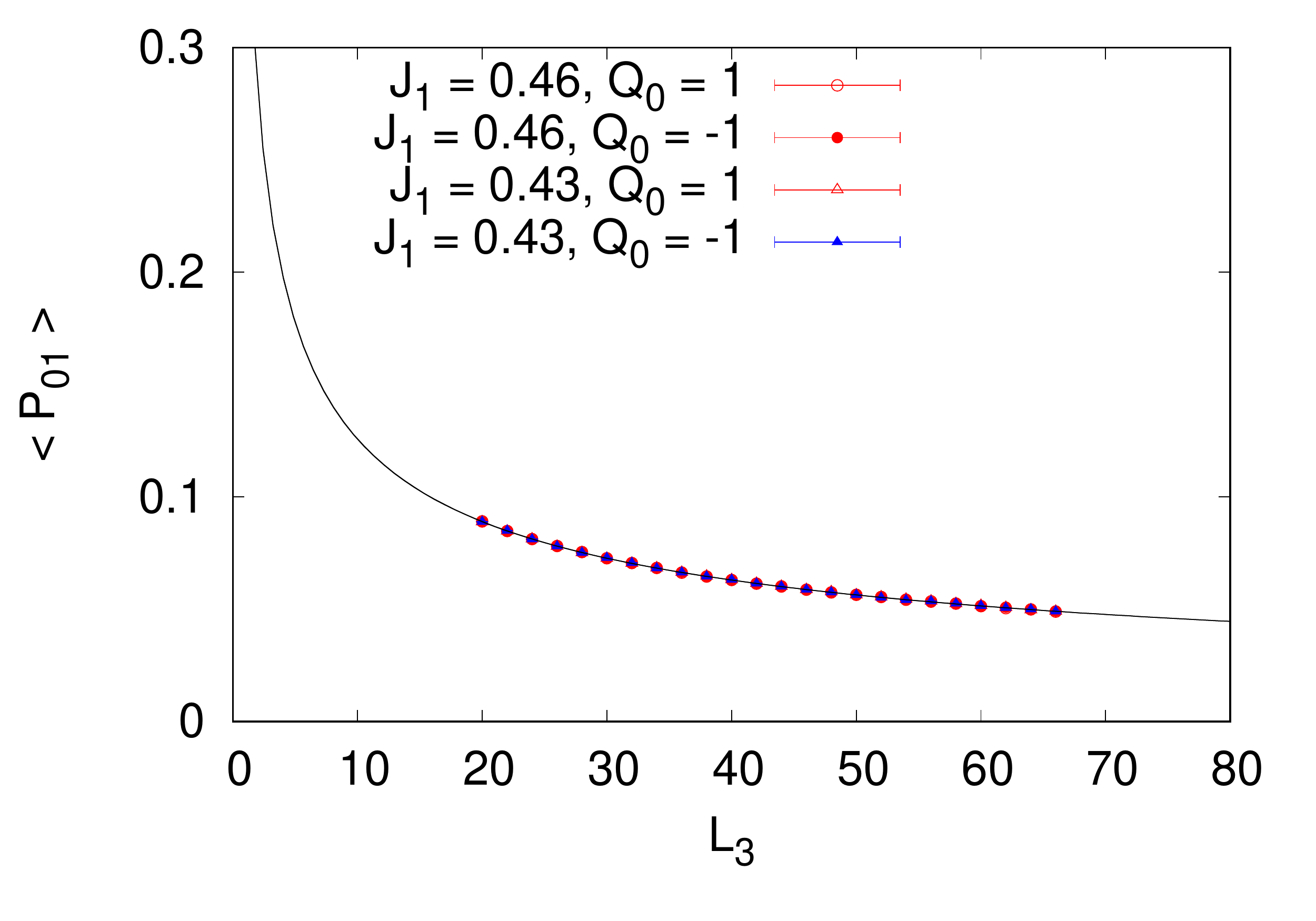}
    \includegraphics[width=0.32\textwidth, angle=270]{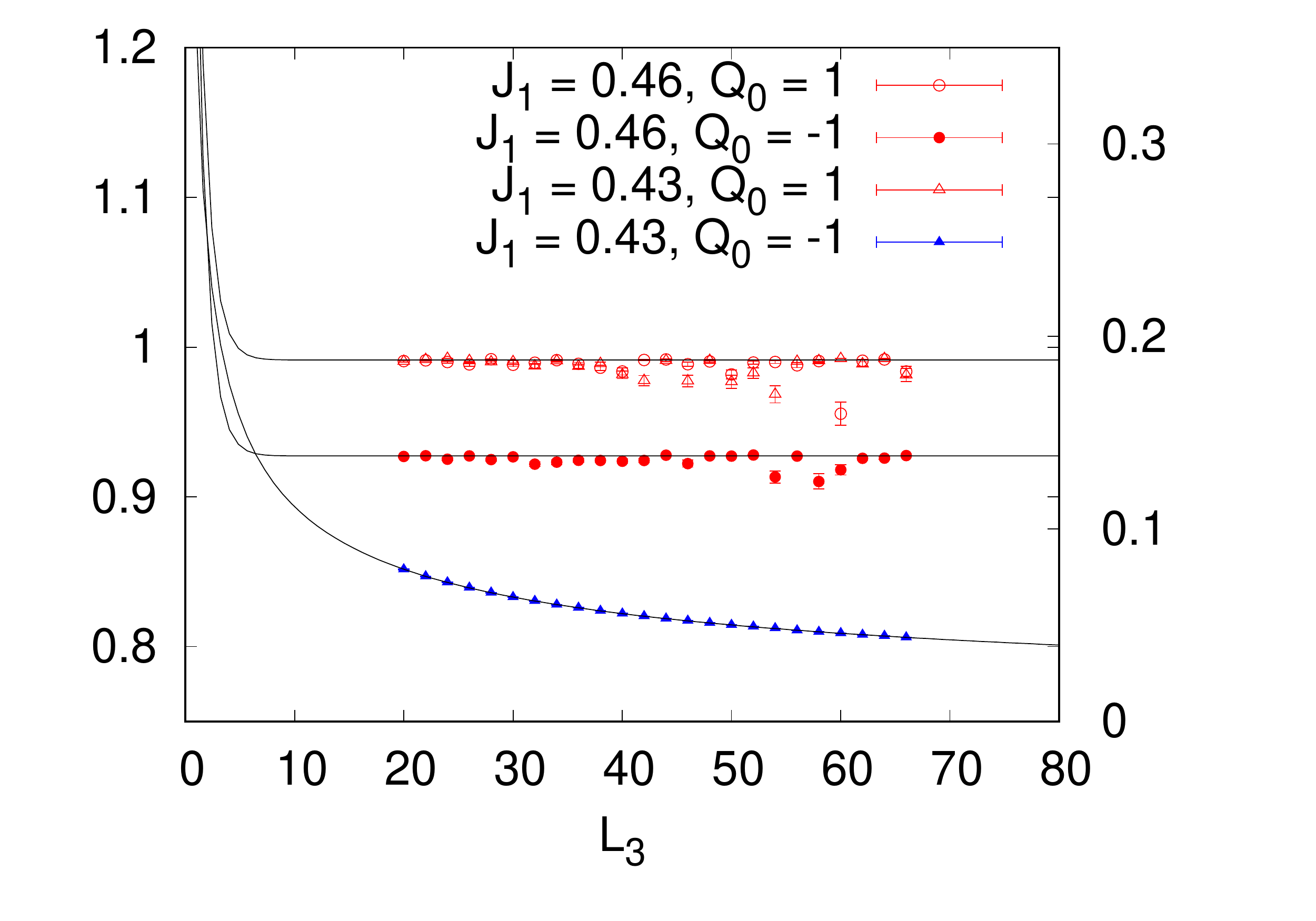}
    \caption{Results for the expectation value of the Polyakov plane}
    \label{fig:plane}
\end{figure}

The last question which we wish to discuss in this contribution is the question whether the numerical estimates provided in Table \ref{tab:polyakov plane values} are not affected by finite volume effects. In principle, it could happen that the non-trivial dependency on $J_1$ and $J_2$ disappears in the infinite volume limit. In order to study this question we performed a series of simulations where we varied the size of the volume perpendicular to the Polyakov line $P_0$ and to the Polykov plane $P_{01}$ by increasing the $L_3$ lattice extend. We present the resulting outcomes in Figure \ref{fig:line} for the Polyakov line and in Figure \ref{fig:plane} for the Polyakov plane. The plots show data at the same four pairs of $(J_1,J_2)$ located in the four regions of the phase space, and again for the two values of the topological charge $Q_0$. Two possible behaviours can be observed: either the data seems constant with increasing $L_3$ with the exception of very small values of $L_3$ where an exponentially suppressed finite volume effect is present, or the data approaches 0 with a $1/\sqrt(L_3)$ behaviour. Note, because the two behaviours affect the expectation values with a vastly separated magnitudes, two vertical scales are used in the figures: data with large expectation value $\approx 0.9$ which does not vanish with increasing $L_3$ has the left vertical scale, whereas the data with small magnitude $\approx 0.1$ which clearly vanish with increasing $L_3$ has the right vertical scale. These figures confirm that the situation described above based on the values in Table \ref{tab:polyakov plane values} is stable as far as the infinite volume is concerned, in the sense that outcomes with small magnitude eventually vanish at large $L_3$ whereas those with large magnitude stay finite and non-zero. This confirms that the Polyakov plane is an interesting observable sensitive to the complex and intertwinned dynamics of both types of dynamical degrees of freedom present in the model. 

\section{Summary and outlook}

In this contribution we gave a brief introduction to 2-group gauge models providing a generalization of gauge theory which includes dynamical higher symmetries. We presented some details of the general construction and their motivations. Subsequently, we specialized to a particular model by choosing a specific 2-group based on two $\mathbb Z_4$ groups and a possible generalization of the Wilson action. We discussed the phase-space of that model, identified two phase transitions and provided order parameters for both of them. For the order parameter for the first-order phase transition in the $J_1$ coupling constant we proposed the Polyakov line, while for the expected second-order phase transition in the $J_2$ coupling constant a new observable: the Polyakov surface. Due to the non-locality of the Polyakov surface, it turned out that it can also serve as an order parameter for the former phase transition. We have performed numerical simulations of the model which confirmed the proposed phase diagram of the model and the expected behaviour of the two order parameters. 

Looking at the broader level, one would like to improve the algorithmic setup in such a way as to include topological charge changes and sample the entire phase space of the model. It would be interesting to check whether other models exist where the factorization theorem does not hold. We keep these research direction for future investigations.

\section*{Acknowledgements}

Computer time allocations 'plgtmdlangevin2' and 'plgnnformontecarlo' on the Prometheus supercomputer hosted by AGH Cyfronet in Krak\'{o}w, Poland was used through the polish PLGRID consortium. B. Ruba acknowledges the support of the SciMat grant U1U/P05/NO/03.39.

\end{document}